\documentclass[sn-mathphys]{sn-jnl}% Math and Physical Sciences Reference Style
\jyear{2022}%

\theoremstyle{thmstyleone}%
\theoremstyle{thmstyletwo}%

\theoremstyle{thmstylethree}%
\usepackage{siunitx}
\usepackage{graphicx}
\graphicspath{{figures/}}

\begin{document}

\title[Article Title]{In-orbit performance of LE onboard \textit{Insight}-HXMT in the first 5 years}

%%=============================================================%%
\author[1]{\fnm{Xiaobo} \sur{Li}}

\author*[1]{\fnm{Yong} \sur{Chen}}\email{ychen@ihep.ac.cn}

\author[1]{\fnm{Liming} \sur{Song}}

\author[1]{\fnm{Weiwei} \sur{Cui}}

\author[1]{\fnm{Wei} \sur{Li}}

\author[1]{\fnm{Juan} \sur{Wang}}

\author[1]{\fnm{Shuang-Nan} \sur{Zhang}}

\author[1]{\fnm{Fangjun} \sur{Lu}}

\author[1]{\fnm{Yupeng} \sur{Xu}}

\author[1]{\fnm{Haisheng} \sur{Zhao}}

\author[1]{\fnm{Mingyu} \sur{Ge}}

\author[1]{\fnm{Youli} \sur{Tuo}}

\author[1]{\fnm{Yusa} \sur{Wang}}

\author[1]{\fnm{Tianxiang} \sur{Chen}}

\author[1]{\fnm{Dawei} \sur{Han}}

\author[1]{\fnm{Jia} \sur{Huo}}

\author[1]{\fnm{Yanji} \sur{Yang}}

\author[1]{\fnm{Maoshun} \sur{Li}}

\author[1]{\fnm{Ziliang} \sur{Zhang}}

\author[1]{\fnm{Yuxuan} \sur{Zhu}}

\author[1]{\fnm{Xiaofan} \sur{Zhao}}

\affil*[1]{\orgdiv{Key Laboratory of Particle Astrophysics}, \orgname{Institute of High Energy Physics, Chinese Academy of Science}, \orgaddress{\street{19B Yuquan Road, Shijingshan District}, \city{Beijing}, \postcode{100049},  \country{China}}} 
%\affil[2]{\orgdiv{Department}, \orgname{Organization}, \orgaddress{\street{Street}, \city{City}, \postcode{10587}, \state{State}, \country{Country}}}

%\affil[3]{\orgdiv{Department}, \orgname{Organization}, \orgaddress{\street{Street}, \city{City}, \postcode{610101}, \state{State}, \country{Country}}}

%%==================================%%
%% sample for unstructured abstract %%
%%==================================%%

\abstract
{\textbf{Purpose:} 
The Low-Energy X-ray telescope (LE) is a main instrument of the \textit{Insight}-HXMT mission and consists of 96 Swept Charge Devices (SCD) covering the 1--10 keV energy band. The energy gain and resolution are continuously calibrated by analysing Cassiopeia A (Cas A) and blank sky data, while the effective areas are also calibrated with the observations of the Crab Nebula. In this paper, we present the evolution of the in-orbit performances of LE in the first 5 years since launch.\\
\textbf{Methods:} 
The \textit{Insight}-HXMT Data Analysis Software package (\texttt{HXMTDAS}) is utilized to extract the spectra of Cas A, blank sky, and Crab Nebula using different Good Time Interval (GTI) selections. We fit a model with a power-law continuum and several Gaussian lines to different ranges of Cas A and blank sky spectra to get peak energies of their lines through \texttt{xspec}.  
After updating the energy gain calibration in CALibration DataBase (\texttt{CALDB}), we rerun the Cas A data to obtain the energy resolution.
An empirical function is used to modify the simulated effective areas so that the background-subtracted spectrum of the Crab Nebula can best match the standard model of the Crab Nebula.\\ 
\textbf{Results:} 
The energy gain, resolution, and effective areas are calibrated every month. The corresponding calibration results are duly updated in CALDB, which can be downloaded and used for the analysis of \textit{Insight}-HXMT data. Simultaneous observations with \textit{NuSTAR} and \textit{NICER} can also be used to verify our derived results.\\
\textbf{Conclusion:} 
LE is a well calibrated X-ray telescope working in 1--10\,keV band.
The uncertainty of LE gain is less than  20\,eV in 2--9\,keV band and the uncertainty of LE resolution is less than 15\,eV. The systematic errors of LE, compared to the model of the Crab Nebula, are lower than 1.5\% in 1--10\,keV.}


%%\pacs[JEL Classification]{D8, H51}

%%\pacs[MSC Classification]{35A01, 65L10, 65L12, 65L20, 65L70}

\maketitle

\section{Introduction}\label{sec1}

The \textit{Insight}-Hard X-ray Modulation Telescope\,(\textit{Insight}-HXMT) was launched on June 15, 2017 at an altitude of 550\,km and an inclination of 43\,degrees \cite{Zhang2020Overview}. It consists of three main telescopes: the High Energy X-ray Telescope (HE, 20--250\,keV, 5000 $\rm cm^{2}$, timing resolution: \SI{2}{\micro\second}) \cite{Liu2020HE}, the Medium Energy X-ray Telescope (ME, 8--35\,keV, 952 $\rm cm^{2}$, timing resolution: \SI{6}{\micro\second})\cite{Cao2020ME}, and the Low Energy X-ray Telescope (LE, 1--10\,keV, 384 $\rm cm^{2}$, timing resolution: \SI{1}{\micro\second})\cite{Chen2020LE}. 
This configuration makes \textit{Insight}-HXMT have broad X-ray energy bands, large areas, and high timing and energy resolution. Therefore, \textit{Insight}-HXMT has the unique ability to study fast X-ray variability in multiple energy bands, allowing it to explore regions closer to black holes or the surfaces of neutron stars than ever before.
As an interesting example, \textit{Insight}-HXMT has identified a non-thermal X-ray burst from SGR J1935+2154, which is associated with FRB 200428 \cite{2021NAlick}, combining the advantages of large areas and broad energy bands.

As a main telescope of \textit{Insight}-HXMT, LE consists of three detector boxes, each containing 32 chips of CCD236 which is a kind of Swept Charge Devices (SCD) \citep{Chen2020LE, 2018SPIElixb}. CCD236, developed by e2v company, has a fast signal readout ability, whereas it loses the position information of photons \cite{2008SPIECCD236}. 
Compared with other types of CCDs, the pile-up effect of LE can be neglected, even when the source flux reaches at about 8 times that of Crab.
Thus LE performs well in studying the timing and energy sepctra of bright sources.
For events with energies above the on-board threshold, their readout time and the ADC channels are measured, digitized, and telemetered to the ground. In addition to triggered events, LE has the forced trigger events, which record the noise amplitude per 32\,ms of each CCD \citep{Li2020InflightCalibration}. The forced trigger events are also saved like triggered events but with a different type and can be used to estimate the noise levels. The \textit{Insight}-HXMT Data Analysis Software package\footnote{http://hxmten.ihep.ac.cn/} (\texttt{HXMTDAS}) has utilized the forced trigger events within 1 second to obtain the average ADC channel of noise, which will be subtracted from the ADC channel of the triggered events in this second. 

The CCD detectors of LE has four quadrants, each with 100*100 pixels.
If a photon interacts with one of these pixels, the resulted charge cloud spreads to the other surrounding pixels, which may induce split events. 
These split events may be read out in adjacent readout periods. The single events without split are considered to calibrate the performance of LE.
The pre-launch calibration experiments and the modeling of the response matrix of LE are performed in \cite{2021JInstzhuyx}. The in-flight calibration method of LE is detailed in \cite{Li2020InflightCalibration} and the calibration of the timing system is reported in \cite{Tuo2022ApJS}. 

In this paper, we focus on the evolution of in-flight performance of LE during the first 5 years of in-orbit operation. Our paper is structured as follows. Section 2 describes the selection of the calibration data and the data reduction processes. 
Section 3 presents the calibration results related to the energy response matrix, including the energy gain, energy resolution, and energy response function. Section 4 describes the calibration of the effective areas based on the Crab Nebula observations. 
%and validates the calibration results using simultaneous observations of \textit{NuSTAR}, \textit{NICER} and \textit{Swift}/XRT. 
Meanwhile, the systematic errors of LE derived using the five-year observations of the Crab Nebula are also given in this section. 
Section 5 summarizes the paper and provides perspectives.

\section{Observations and data reduction}\label{sec2}

\subsection{Data reduction for energy gain calibration} \label{sec2.1}
\begin{figure}[H]%
    \centering
    \includegraphics[width=0.9\textwidth]{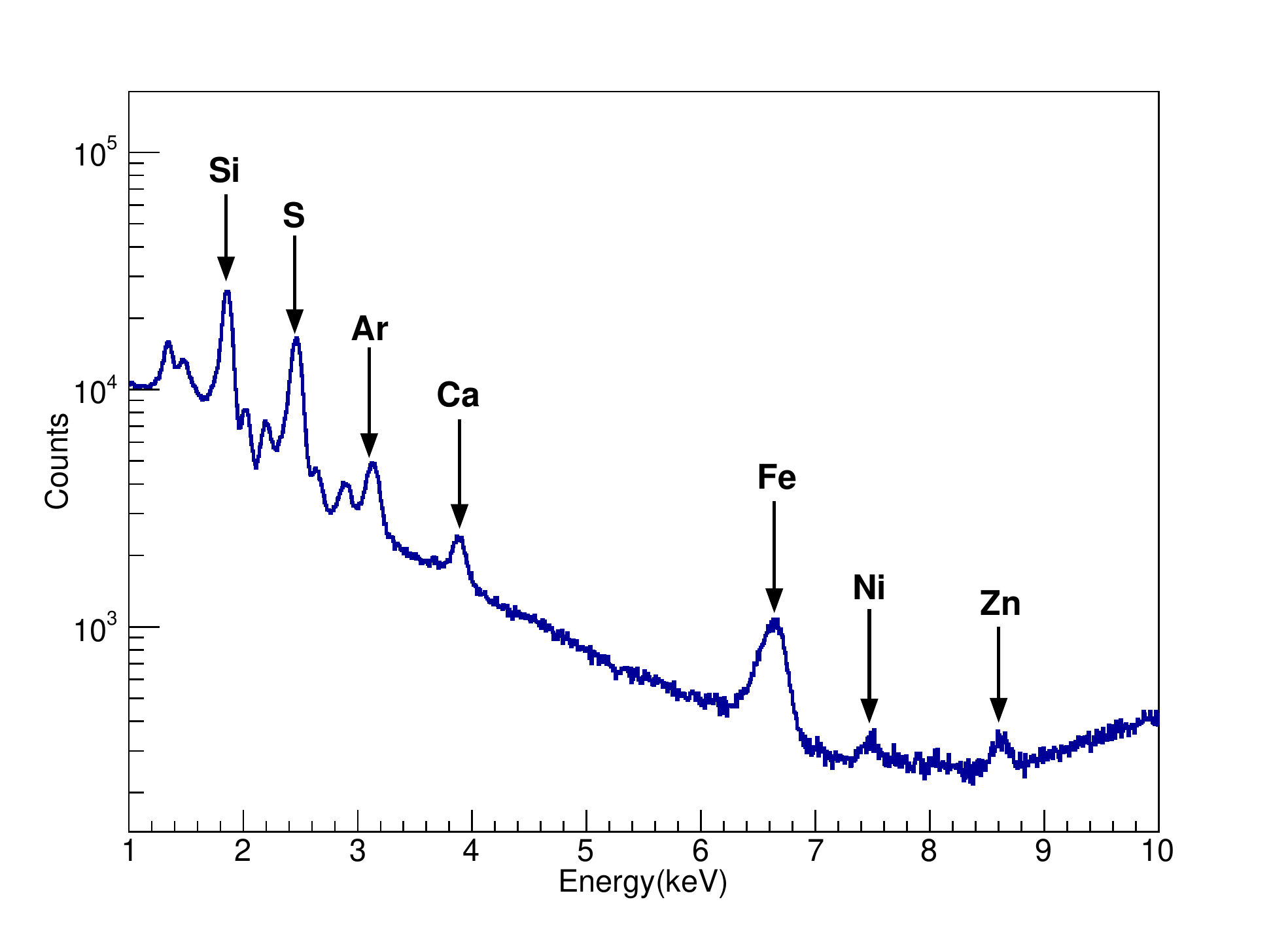}
    \caption{The energy spectrum of Cas A observed on 2017-07-08. The rich emission lines of Cas A (Si, S, Ar, etc) and some background lines of Ni and Zn are marked with black arrows. These emission lines can be used to calibrate the gain of LE.} \label{figcasApeak}
\end{figure}

To investigate the energy gain and resolution calibration, the data suitable for parameterizing the energy gain and monitoring its variations come from the sources that produce visible peaks in the observed spectra, such as blank sky observations with internal background X-ray florescence lines (Ni, Cu, and Zn) and the supernova remnant Cassiopeia A (Cas A) with rich emission lines (Si, S, Fe, etc). The Cas A spectrum observed on July 8, 2017 is displayed in Figure \ref{figcasApeak}. The prominent emission lines can be seen clearly and used to calibrate the energy gain and resolution.

Cas A is observed as a calibration source almost every month if the solar angle is allowed. It should be noted that Cas A is not visible to \textit{Insight}-HXMT from March to June of each year. The calibration results for these months should be extrapolated from the data before and after these months.
The emission lines of Ni, Cu, and Zn produced by materials near the CCD detectors during the blank sky observations can also be used to calibrate the energy gain of LE. 
From 2017-07-08T01:02:26 \footnote{The time format (YYYY-MM-DDTHH:MM:SS) used in the paper is UTC time.} to 2022-09-27T03:01:53, the data of pointing observations of Cas A (ObsID P0101326, P0202041, P0302291, P0402348, P0502131) are reduced to monitor and calibrate the gain in 1.3--6.7\,keV. In addition, from 2017-11-02T05:00:54 to 2022-09-29T23:17:57, the data of pointing observations of the blank sky (ObsID P0101293, P0202041, P0301293, P0401293, P0501293) are also reduced to calibrate the gain in 7.4--8.6\,keV.

\texttt{HXMTDAS} and the pre-launch CALibration DataBase (\texttt{CALDB}) are used to reduce and calibrate the data. Here, step-by-step commands are presented as follows.

\begin{itemize}
  \item  \texttt{lepical} is used to calibrate the photon events of LE using the pre-launch Energy-Channel (E-C) relation. Meanwhile, the noise events per second are accumulated to obtain the noise level.
  \item \texttt{lerecon} is applied to the calibrated events of LE to identify the single or split events and reconstruct split events into one event.
  \item The Good Time Interval (GTI) can be calculated using  \texttt{legtigen}, with the criterion \texttt{ELV>10 \&\& ANG\_DIST<=0.04 \&\& COR>8 \&\& SAA\_FLAG==0 \&\& T\_SAA>=100 \&\& TN\_SAA>=100}. In addition, the calibrated event data can be filtered using the GTI through \texttt{lescreen}.
  \item  Finally, the Cas A spectra of single events can be generated from \texttt{lespecgen}.
  \end{itemize}

After the reduction of \texttt{HXMTDAS}, we fit a model with a power-law continuum and several Gaussian lines to different ranges of Cas A spectra using \texttt{xspec}. The peak energies of Cas A emission lines using a pre-launch E-C can then be obtained.
The energy peaks of background lines can also be derived when we use the same procedure as mentioned above to reduce the blank sky data.
Consequently, the energy evolution of different lines versus time can be obtained using the same E-C.

\subsection{Data reduction for energy resolution calibration} \label{sec2.2}
After calibrating the gain of LE, we put the new gain file into CALDB and apply the updated CALDB to generate the Cas A spectra again.
The widths of Si, S, and Fe using the pre-launch response file of LE are jointly fitted with \emph{XMM/MOS} (observed on June 22, 2006 with obsID 0412180101) in \texttt{Xspec}. If the energy resolution of LE keeps the same as the ground-based measurements, the intrinsic width of Si, S, and Fe will be the same as the fit result of \emph{XMM/MOS}.
Actually, the fitted intrinsic widths of Si, S, and Fe for LE are larger than those for \emph{XMM/MOS}. Therefore, the energy resolution also changes compared with the pre-launch calibration results.
After subtracting the intrinsic width of Si, S, and Fe from the fitted results of \emph{XMM/MOS}, an additional broadening of LE resolution is obtained, which evolves with time and temperature as described in \cite{Li2020InflightCalibration}.

\subsection{Data reduction for effective areas calibration} \label{sec2.3}
The Crab Nebula (together with its pulsar) that originated from a supernova explosion in 1054 AD is well known as the standard candle in the X-ray sky due to its brightness, almost constant intensity, and simple power-law distribution of emission spectra in the band from 1 to 100\,keV. Therefore, many X-ray astronomy satellites perform their calibration with the Crab Nebula \citep{Madsen2015Crab, Tuo2022ApJS}. Here, the data of \textit{Insight}-HXMT's pointed observations of the Crab Nebula from 2017-08-27T04:05:29 to 2022-09-05T03:43:48 (ObsID P0101299, P0111605, P0202041, P0303390, P0402349, P0502132) are reduced to calibrate the effective areas. 

Once the gain, resolution, and response matrix file are updated in the CALDB, we will use \texttt{HXMTDAS} and the latest CALDB to extract the Crab Nebula spectra including the contribution of background. The difference with the reduced Cas A and blank sky is the GTI criterion, i.e.: \texttt{ELV>10 \&\& ANG\_DIST<=0.04 \&\& COR>8 \&\& SAA\_FLAG==0 \&\& T\_SAA>=300 \&\& TN\_SAA>=300}. This criterion consists of two parts; the first part \texttt{ELV>10 \&\& ANG\_DIST<=0.04} is used to select the Crab Nebula in the field of view, while the second part \texttt{COR>8 \&\& SAA\_FLAG==0 \&\& T\_SAA>=300 \&\& TN\_SAA>=300} should ensure that the LE background model is available\citep{Liao2020LEBackground}. Furthermore, we use \texttt{legticor} to improve the accuracy of the LE background estimation. Finally, \texttt{lebkgmap} is also applied to generate the background spectra in the GTI. 

\section{Calibration results of energy response matrix}\label{sec3}
\subsection{The performance of noise} \label{sec3.1}
In order to study the background model, blank sky observations are performed almost every month.
Consequently, the forced trigger events of blank sky observations can be used to monitor the gain and resolution of each CCD. 
To obtain the average peak and full width at half maximum (FWHM) of each CCD, we use a Gaussian function to fit the Channel distribution of the forced trigger events per minute. The mean and the sigma of the Gaussian function represent the peak and width of the noise events in this minute, respectively.

\begin{figure}[H]%
\centering
    \includegraphics[width=0.5\textwidth]{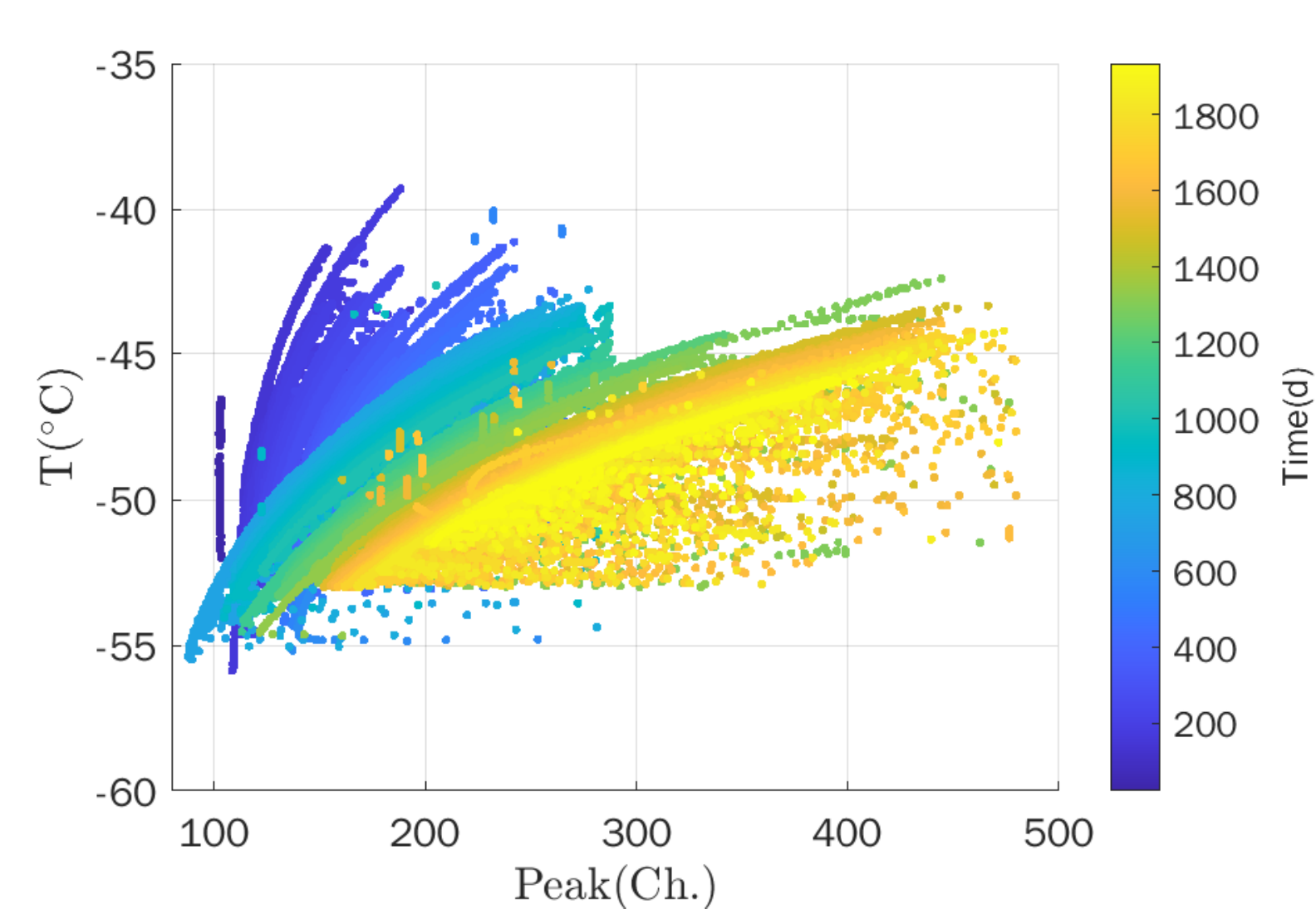}
    \includegraphics[width=0.45\textwidth]{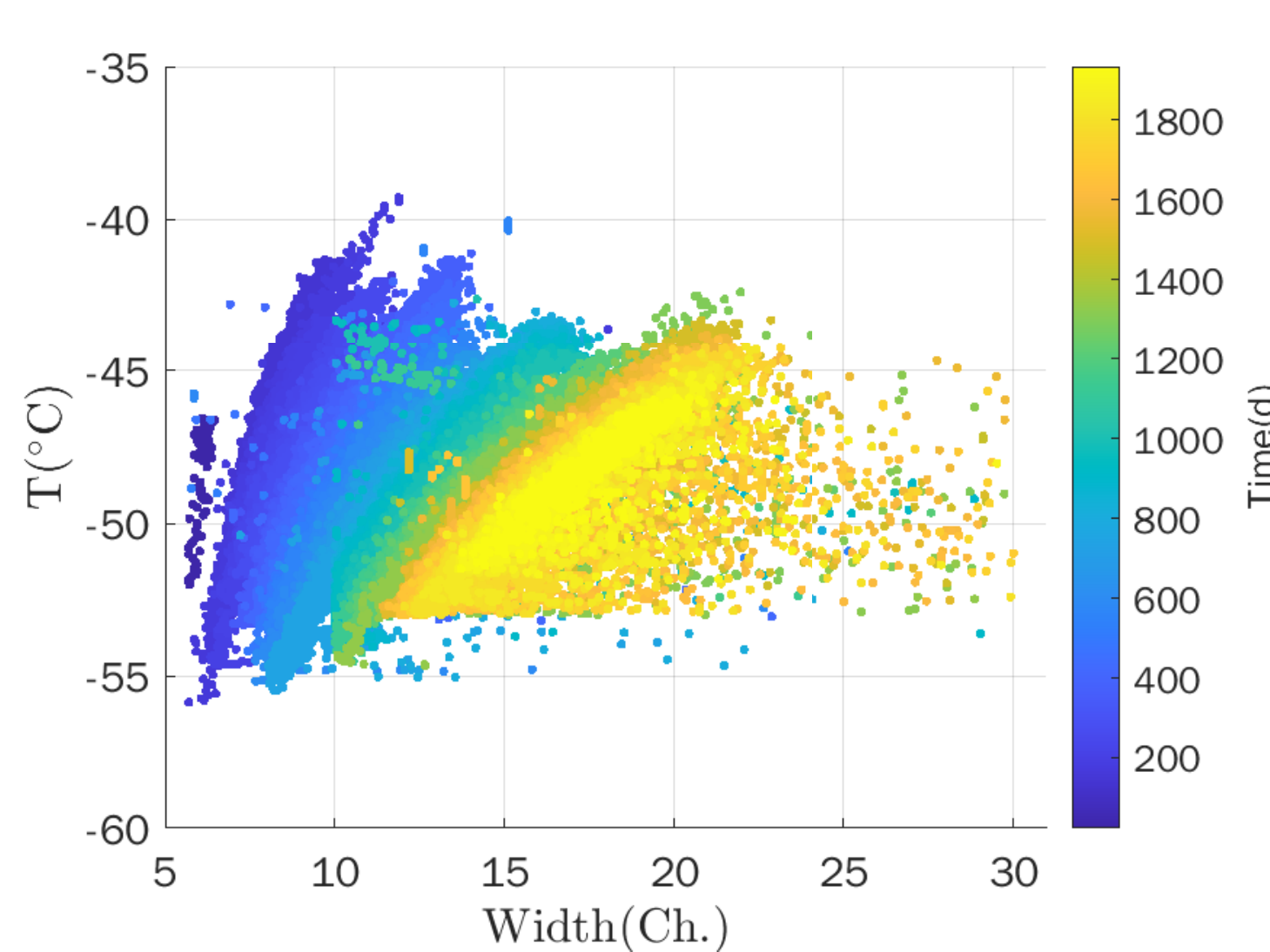}
\caption{Left: The peak of forced trigger events versus temperature and time. Right: The width of forced trigger events versus temperature and time. The colors on the right of each panel represent the days since launch. }
\label{figpeakwidth}
\end{figure}

In Figure \ref{figpeakwidth}, we show the fitting results of peak and width of forced trigger events, and also calculate and display the temperature and the number of days since launch for those forced trigger events. The colors represent the number of days since launch. The peak increases with increasing temperature. The peak changes more and more as time increases. For example, since the time is less than 200 days, the change starts with less than 50 channels, and for days over 1500 days it has a maximum change of about 250 channels.

%In Figure \ref{figpeakwidth}, the fitting results shows the evolution over time between peak (width) and temperature of a CCD and the colors represents the days since launch. The peak increases with increasing temperature. As time increases, the peak changes more and more. For example, the change starts with less than 50 channels and now has a maximum change of about 250 channels.
The phenomenon is similar to the width, which also becomes larger as the temperature and time increase.
This indicates that the radiation in space causes an increase in the noise level.

Due to the limited resources on the satellite, LE adopted a passive radiation cooling technique to control the temperature of the detector. The stability of the temperature control is fairly good for LE. It should be noted that the in-orbit working temperature of LE is measured every second for each CCD, with a temperature variation of about 5 degrees per orbit. From the LE five-year data,  the temperature of all CCDs varied between $-55^{\circ}\mathrm{C}$ and $-38^{\circ}\mathrm{C}$, with an average of $-47.9^{\circ}\mathrm{C}$ and a step of $0.01^{\circ}\mathrm{C}$.
In the data reduction of \texttt{HXMTDAS}, \texttt{lepical} and \texttt{lerspgen} all require temperature files to retrieve the correct calibration files and parameters in CALDB.

\subsection{Energy gain} \label{sec3.2}
\begin{figure}[H]%
    \centering
    \includegraphics[width=0.85\textwidth]{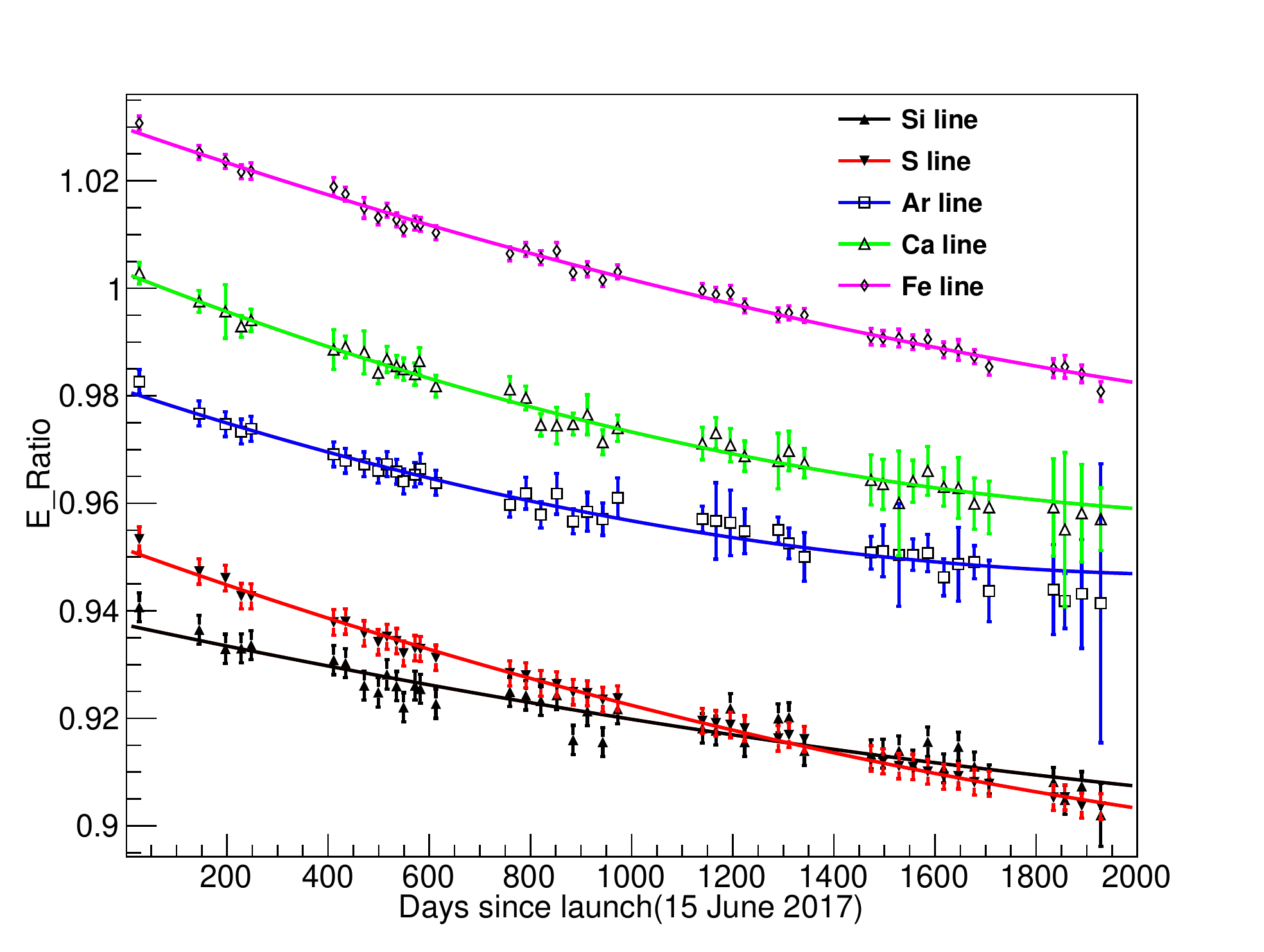}
    \caption{Energy fit (with 1$\sigma$ error bars) of different emission lines against time in Cas A spectra. The peak of the emission lines obtained by the pre-launch E-C are divided by the theoretical energy of that line and then the ratios of the fitted energy to the model energy are scaled by different constants. The peak energies decrease with time and a simple quadratic polynomial fit is used to describe the evolution.} \label{fig2}
\end{figure}

\begin{figure}[H]%
    \centering
    \includegraphics[width=0.8\textwidth]{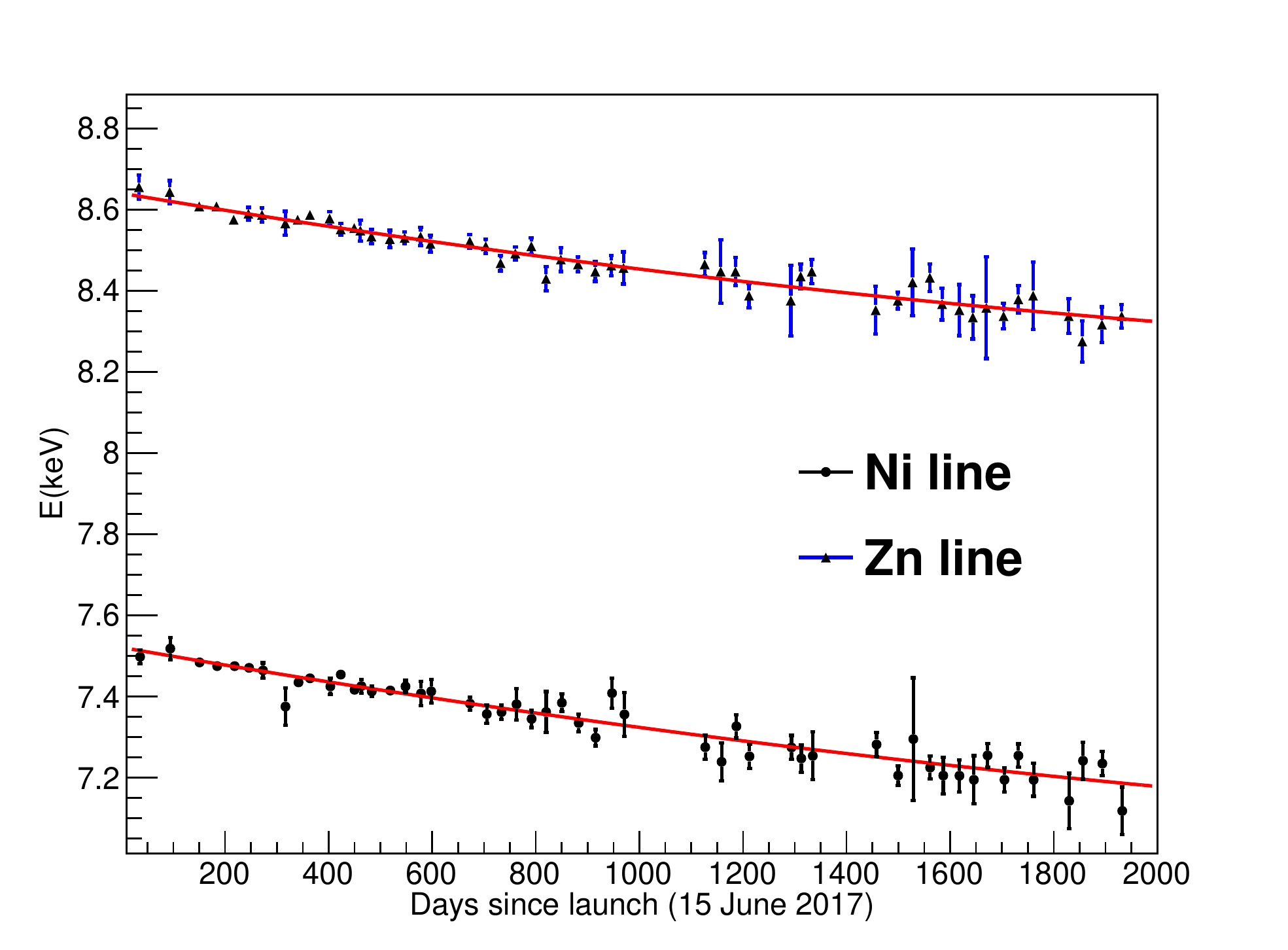}
    \caption{Energy fit (with 1$\sigma$ error bars) of Ni and Zn versus time in blank sky spectra. The peaks of the lines also decrease with time and also a simple quadratic polynomial fit is used to describe the evolution as Figure \ref{fig2}.} \label{fig3}
\end{figure}

We use the pre-launch E-C results stored in CALDB at different temperatures to fit the peak energies of the emission lines depicted in Figure \ref{figcasApeak} with all the data from Cas A and blank sky.
To clearly show the evolution of Cas A emission lines, all the peak energies are divided by the theoretical energy of that line and then the ratios of the fitted energy to the model energy are scaled by different constants as shown in Figure \ref{fig2}.  The energy fit results of Ni and Zn in blank sky spectra are also displayed in Figure \ref{fig3}.

As noticed from Figures \ref{fig2} and \ref{fig3}, peak energies of emission lines detected by LE from Cas A and blank sky are gradually decreasing.
This phenomenon may be caused by the decrease of the charge transfer efficiency of the CCDs due to radiation damage in space.
In order to describe the evolution, a quadratic polynomial function is used to fit the change. From the results of fit, the peak values can be derived on any day, even when Cas A is not visible due to the observation limitations.
Finally, the new E-C parameters can be obtained from the lines above and the detailed description of the method can be found in literature \cite{Li2020InflightCalibration}.

\subsection{Energy resolution} \label{sec3.3}
Once the gain file has been updated into CALDB, we rerun the Cas A data to get the gain corrected spectra.
The widths of Si, S, and Fe using the pre-launch response file of LE are jointly fitted with \emph{XMM/MOS} in \texttt{Xspec}. Then the additional broadening of Si, S, and Fe for different observations can be derived after subtracting their intrinsic width from the fitted results of \emph{XMM/MOS}. 

\begin{figure}[H]%
    \centering
    \includegraphics[width=1.0\textwidth]{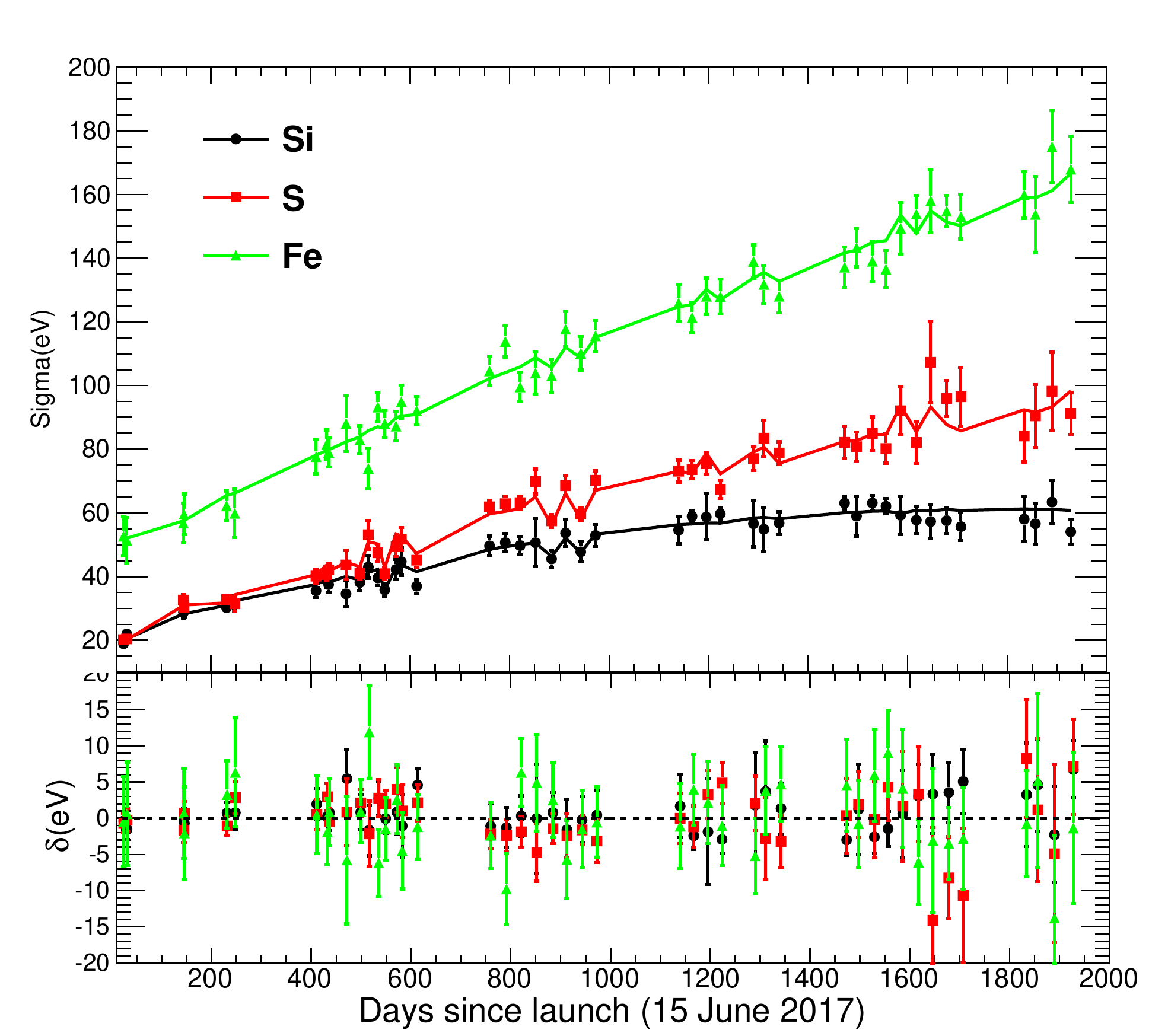}
    \caption{Additional broadening of the width sof Si, S, and Fe plotted against the observation time at different temperature. The model described in \cite{Li2020InflightCalibration} is also plotted as the solid lines. In the lower panel, the residuals of the fit are plotted against time for the Si, S, and Fe respectively. The differences between data and model are less than 15\,eV.} \label{fig4}
\end{figure}

For an observation of Cas A, we calculate the average temperature in its GTI and the observation time with the weight of its exposure time. The additional broadening width are plotted against the observation time at different temperatures for Si, S, and Fe line in Figure \ref{fig4}. 
Using the two-dimensional Equation (6) in literature \cite{Li2020InflightCalibration} to fit the additional broadening of LE, we can obtain the additional broadening at any time even when the Cas A is not visible.
The residuals are plotted in the bottom panel of Figure \ref{fig4}. 
The differences between data and model are less than 15\,eV.
Therefore, the two-dimensional function describes the large fluctuations caused by temperature in the top panel of Figure \ref{fig4} very well.
Then a linear function is used to fit the extra broadening of LE versus energy.

\subsection{Energy response function} \label{sec3.4}
Since LE is a CCD detector, the diffusion of the charge cloud generated by a photon or a charged particle over several pixels can be read out by several adjacent cycles. Here, we only consider single events without splitting as valid X-ray events.

\begin{figure}[H]%
    \centering
    \includegraphics[width=0.8\textwidth]{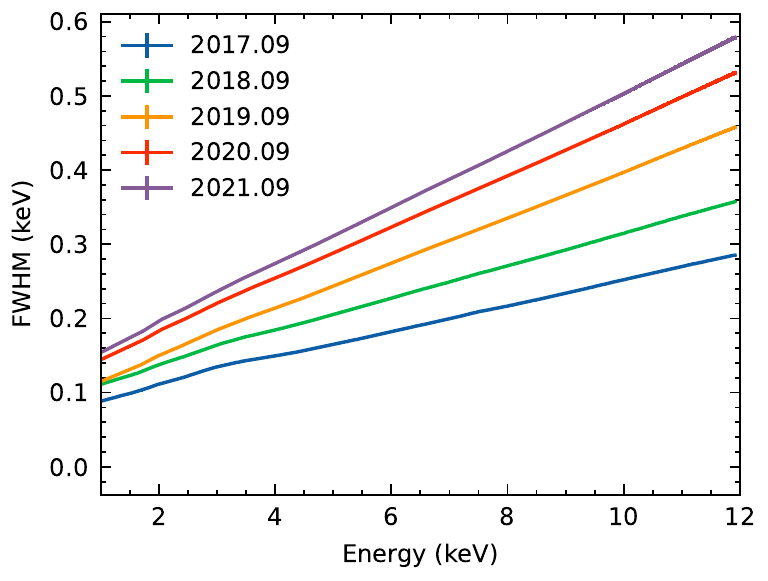}
    \caption{The FWHM of LE versus energy for different observation years in September. The FWHM has been increasing, but the rate of increase has decreased recently.} \label{figrmf}
\end{figure}

The pre-launch Response Matrix File (RMF) was calibrated at the calibration facility of the Institute of High Energy Physics, using 20 discrete energies covering the energy range 0.9--12\,keV \cite{2021JInstzhuyx}.
When we use the pre-launch RMF to fit the in-flight spectrum of Cas A, additional broadening is needed to fit the line profiles well, as described in section \ref{sec3.3}. 
To generate the in-flight RMF, the additional broadening versus energy requires to be convolved with the pre-launch RMF.
After convolution, we use a Gaussian function to fit the full energy peak to obtain the FWHM. The FWHM versus energy for different observation year is plotted in Figure \ref{figrmf}.
As pictured, the FWHM has been increasing, but the rate of increase has decreased in the last two years.

\section{Calibration results of effective areas and systematic errors}\label{sec4}

%\subsection{Split ratio}

\subsection{Effective areas}
The spectrum of the Crab Nebula in the 1--100\,keV X-ray band has been well described by a power-law with photon index of $\Gamma \sim 2.11$ \citep{2015ApJSNuSTAR}.
As for the normalization factor, the Crab spectrum measured by \emph{NuSTAR} in March, 2018 is fitted with $N = 8.76\,{\rm {keV^{-1}cm^{-2}s^{-1}}}$.
The Crab Nebula is modeled as a simple absorbed power law,
\begin{equation}\label{equ:crabmodel}
   F(E)=\rm{abs}(\emph{E})\emph{N}\emph{E}^{-\Gamma},
\end{equation}
where $\rm{abs}$ is the interstellar absorption model, \emph{N} is the normalization factor, $\Gamma$ is the power-law photon index, and $E$ is the photon energy. We define the Crab model with $\Gamma=2.11$, ${N}=8.76\,{\rm {keV^{-1}cm^{-2}s^{-1}}}$ and $\mathrm{nH}=0.36\times10^{22}\,{\rm cm^{-2}}$ which is a parameter in $\rm{abs}$.

\begin{figure}[H]%
    \centering
    \includegraphics[width=1.0\textwidth]{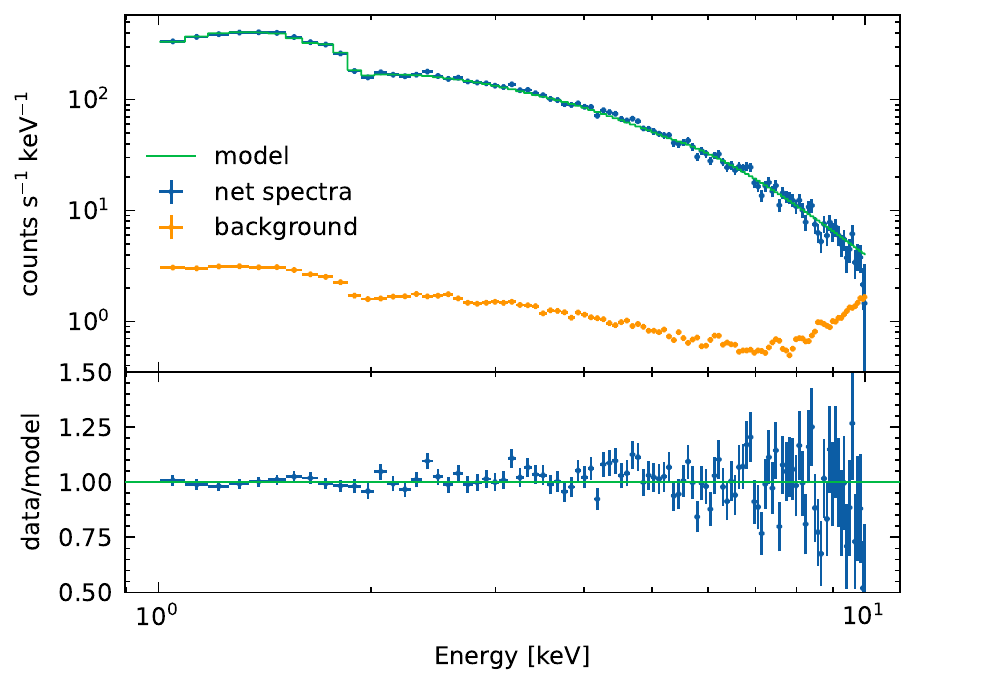}
    \caption{The top panel shows the energy spectrum of the Crab Nebula observed during a GTI (from 2018-02-08T16:28:45 to 2018-02-08T16:29:42, obsID: P0111605046). The green line is the model described in Equation (\ref{equ:crabmodel}) with the parameters of $\mathrm{nH}=0.36\times10^{22}\,{\rm cm^{-2}}$ and $\Gamma=2.11$. The data points in blue are background-subtracted count rate, while the data points in orange are the corresponding background count rate estimated from \texttt{lebkgmap}. For clarity, these data points are binned every 30 PI channels. The bottom panel presents the ratio of data (blue points in the top panel) to model (green lines in the top panel).} \label{figcrabspec}
\end{figure}

For the observed spectrum of Crab Nebula in GTI, the detected rate in a given instrumental pulse invariant (PI) bin, $S(PI)$, can be modeled as the equation,
  \begin{equation}\label{equ:EffArea}
   S(PI)=F(E)\times A(E)\ast RMF(PI, E) + B(PI),
 \end{equation}
where $F(E)$ is the model photon spectrum of the Crab as a function of the incident photon energy, $RMF(PI, E)$ is the redistribution matrix that represents the probability density in a given PI bin for the photons with energy $E$, $A(E)$ is the effective area, also known as the ancillary response function (ARF), and $B(PI)$ is the background count rate in the GTI. In the top panel of Figure \ref{figcrabspec}, we show the background-subtracted count rate of Crab Nebula in blue points observed on February 8, 2018 during a GTI (from 2018-02-08T16:28:45 to 2018-02-08T16:29:42). The corresponding background count rate estimated from \texttt{lebkgmap} is shown in orange points and the green line represents the model of Crab Nebula described in Equation (\ref{equ:crabmodel}) with the parameters of $\mathrm{nH}=0.36\times10^{22}\,{\rm cm^{-2}}$ and $\Gamma=2.11$.

\begin{figure}[H]%
    \centering
    \includegraphics[width=0.8\textwidth]{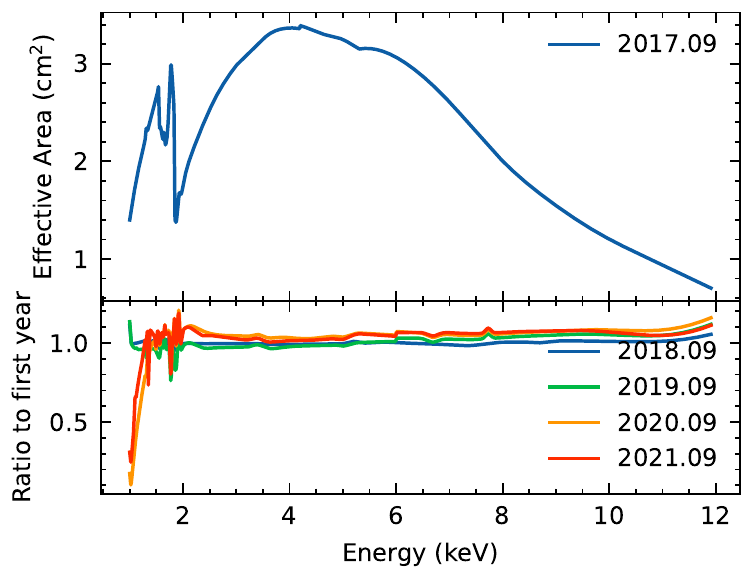}
    \caption{The top panel shows the effective areas of one CCD as a function of energy in September 2017. 
    The bottom panel shows the ratios to the effective areas in 2017. Some large differences can be found after June 2020.} \label{figeff}
\end{figure}

Prior to launch, we used ground calibration results and Monte Carlo simulations based on Geant4 toolkit to produce the effective areas. After launch, there still remain systematic residuals in the Crab spectrum using the new simulated effective areas as shown in Figure 23 of \cite{Li2020InflightCalibration}. 
Although it is desirable to have a fully physics-based effective areas, this is not often achievable with limited number of calibration sources. Moreover, the parameters in the charge transfer process is difficult to quantify accurately.
Finally, it is decided to use an empirical function $f(E)$ to modify the simulated effective areas. Since $f(E)$ is a function of energy $E$, its effect should be folded through the response matrix.
We optimize the empirical function and make the residuals within an acceptable level as shown in the bottom panel of Figure \ref{figcrabspec}. The parameters of the empirical function can be derived and the effective areas in orbit can be represented as $f(E)*A(E)$.
Further details about how we obtain the in-orbit effective areas are available in \cite{Li2020InflightCalibration}.
Since the number of bad CCDs kept changing over the last 5 years as shown in Table \ref{tab:badccd}, here we only show the effective areas of one CCD in September 2017 in the top panel of Figure \ref{figeff}.
The ratios to the effective areas in 2017 are also plotted at the bottom of Figure \ref{figeff}.

\begin{table}[htb]\footnotesize
  \centering
  \caption{Bad CCDs of LE. UTC time when CCDs become not working and their FoVs.}\label{tab:badccd}
   \begin{tabular}{ccc}
    \hline
    detID  & Time (UTC)  & FoV \\
    \hline
     29  & 2017-06-24T15:37:26 & small \\
    \hline
     87  & 2017-06-24T15:37:26 & small \\
   \hline
      1  & 2017-07-28T20:00:30 &  big  \\
    \hline
      15 & 2017-09-15T06:06:55 &  big   \\
    \hline
     69 & 2019-04-14T19:36:29 & big \\
   \hline
     54 & 2019-06-18T15:44:11 & small \\
    \hline
     76 & 2020-06-30T19:42:13 & small \\
    \hline
\end{tabular}
\end{table}

\begin{figure}[H]%
    \centering
    \includegraphics[width=0.8\textwidth]{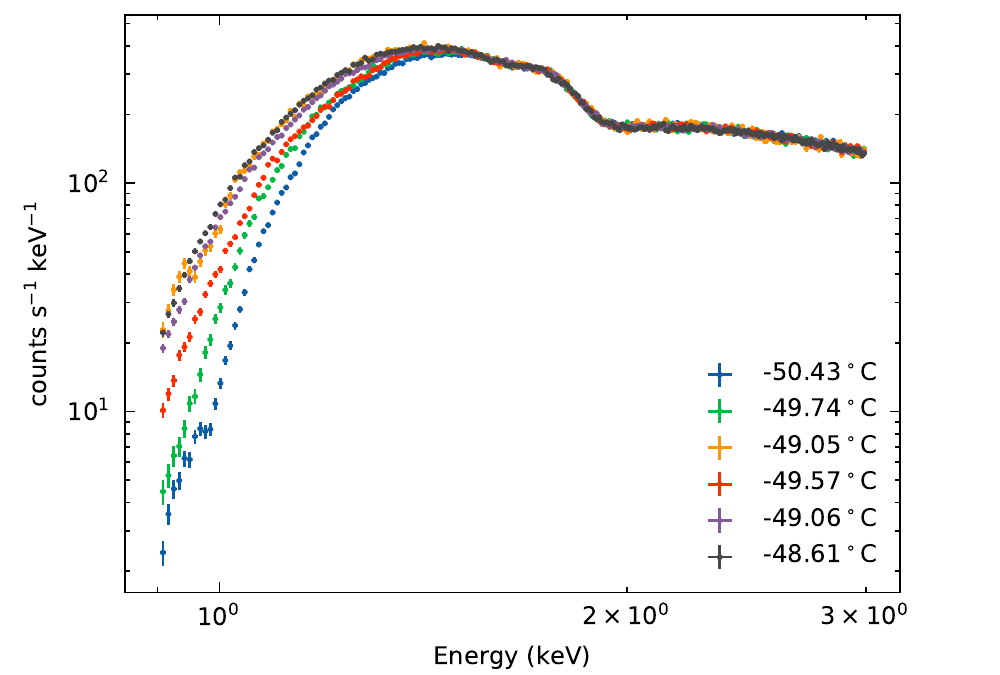}
    \caption{The energy spectra of Crab Nebula observed in 0.9--3\,keV from 2020-09-13T00:40:59 to 2020-09-14T00:50:51 (obsID P0302290002). Six GTIs are available using the standard data reduction procedures. The data points of different colors represent the Crab Nebula spectra in different GTIs and the average temperatures in these GTIs are also calculated and shown. The count rates in 1--1.7\,keV are significantly lower than that in Figure \ref{figcrabspec}. Moreover, the higher the temperature, the higher the count rate.} \label{figcrabtemp}
\end{figure}

It is worth noting that some large differences of effective areas appear below 2\,keV after June 25, 2020. 
This phenomenon can be explained as follows. In order to reduce the rate of noise events, the thresholds of most CCDs are adjusted at about ~450 channel. As pictured in Figure \ref{figpeakwidth}, the X-ray events between 1\,keV and 1.7\,keV have a variation of about 200 channels at different temperatures like the forced trigger events. This will result in some X-ray events exceeding the threshold and others not exceeding it. So the detected count rate has decreased compared to the results before June 25, 2020 as shown in the top panel of Figure \ref{figcrabspec}.
Figure \ref{figcrabtemp} shows the energy spectra of Crab Nebula in 0.9--3\,keV energy band observed from 2020-09-13T00:40:59 to 2020-09-14T00:50:51 (obsID P0302290002). 
Six GTIs can be obtained using the standard data reduction procedures of \texttt{HXMTDAS}. The data points of different colors represent the Crab Nebula spectra in different GTIs and the average temperatures in these GTIs are also shown. The count rates in 1--1.7\,keV have a strong positive correlation with temperature because higher temperatures make more events over the threshold.
Therefore, the efficiency below 1.7\,keV varies rapidly with temperature and decreases significantly with decreasing temperature. It is difficult to calibrate the efficiency well here, so we recommend users to use energy band in 2 - 10 \,keV for spectral analysis when analyzing data observed after June 25, 2020.

After the in-flight effective areas are calibrated, some simultaneous observations with \emph{NuSTAR} and \textit{NICER} are used to validate all calibration results.
An analysis of joint spectra from \textit{Insight}-HXMT, \textit{NuSTAR}, and \textit{INTEGRAL} was performed in
\cite{2022ApJZdziarski}. The authors found an overall agreement between the spectra from all three satellites.
The literature \cite{2022MNRASwangpj} also investigated the 2018 outburst of the black hole transient H 1743 - 322 with a series of \textit{Insight}-HXMT, \textit{NICER}, and \textit{NuSTAR}, covering the 1--120 keV band. They also jointly fitted the spectra of the H 1743 - 322 and obtained consistent results.

\subsection{Systematic errors}
After the effective area files are updated in CALDB, we have reprocessed all Crab data and generated the corresponding response files and background files for LE at different observation times.
The information of all the Crab Nebula data analyzed in this paper is summarized in Table \ref{tab:tab1}.
The first column represents the period of observation and the second column gives the corresponding observation ID. The third column provides the number of exposures in this observation period. 
Typically, an observation lasts several hours or days and the amount of data exceeds several GB. 
To reduce the size of a single file, an observation is artificially split into multiple segments (named exposures). The duration of an exposure is usually about three hours.
The last column provides the average effective exposure time for all exposures in this observation period.

\begin{table}[htb]\footnotesize
  \centering
  \caption{Summary of Crab Nebula data analyzed in this paper.}\label{tab:tab1}
   \begin{tabular}{|c|c|c|c|}
    \hline
    Observation Period & ObsID & Exposure Amount & Mean Exposure \\
    & & & Time (s) \\
    \hline
     2017.08--2018.08 & P0101297,P0111605, P0101299 &  166  &  1820 \\
    \hline
    2018.09--2020.04 & P0101299, P0202041 & 135   & 1421  \\
   \hline
    2020.08--2022.04 & P030229, P0402349 & 129 &  1329  \\
   \hline
\end{tabular}
\end{table}

\begin{figure}[H]%
    \centering
    \includegraphics[width=0.8\textwidth]{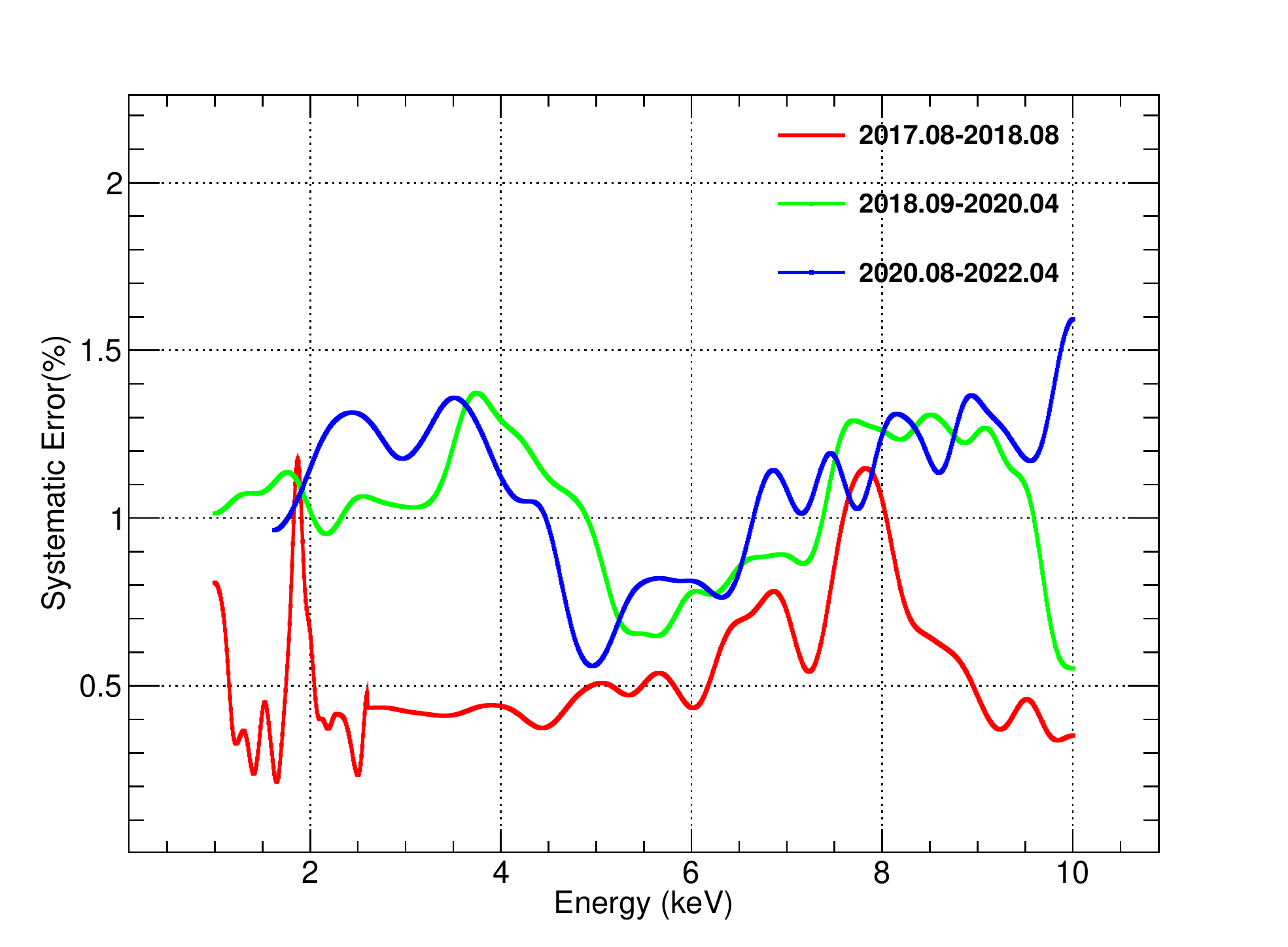}
    \caption{The systematic errors of LE versus energies for different times as shown in different colors. These values can be used for the spectral fitting of LE if the systematic error dominates over statistical errors in the observations.} \label{figsyserror}
\end{figure}

The model of Crab Nebula can be fixed to obtain the ratio of each individual exposure in each PI channel as displayed in the bottom panel of Figure \ref{figcrabspec}.
We can calculate the systematic errors of the ratio at each PI channel using the same method as described in Equation (11) of \cite{Li2020InflightCalibration}. 
In our analysis, only the exposures with an effective exposure time of more than 1000\,s are considered in the calculation of systematic errors.
Figure \ref{figsyserror} depicts the systematic errors versus energies for different times, which are larger after September 2018.
Compared to the model of the Crab Nebula, the systematic errors of LE are below 1.5\% in 1--10\,keV.
These values can be used for the spectral fitting of LE if the systematic error dominates over statistical errors when the source is bright.

\section{Conclusion}\label{sec5}  
LE is a well-calibrated X-ray astronomical satellite working in 1--10\,keV.
The uncertainty of LE gain is less than 20\,eV in 2--9\,keV band and the uncertainty of LE resolution is lower than 15\,eV. 
The effective areas are calibrated with the Crab Nebula using a simple absorbed power law model with photon index $\Gamma =2.11$, normalization factor $N = 8.76\,{\rm {keV^{-1}cm^{-2}s^{-1}}}$, and interstellar absorption $\rm{nH} = 0.36\times10^{22}\,{\rm cm^{-2}}$.
The systematic errors in the spectral fitting are lower than 1.5\% in 1--10\,keV and slightly higher above 6\,keV.
The calibrations are made monthly and updated appropriately in CALDB, which can be downloaded and used to analyze \textit{Insight}-HXMT data.
We recommend that users utilize energy bands from 2 to 10 keV when making spectral analysis for observations after June 25, 2020.
We plan to continue improving the spectral capabilities of the LE and monitoring the detector gain, resolution and effective areas.

\backmatter
\bmhead{Acknowledgments}

This work used data from the \textit{Insight}-HXMT mission, a project funded by the China National Space Administration (CNSA) and the Chinese Academy of Sciences (CAS).  We gratefully acknowledge the support from the National Program on Key Research and Development Project (Grant No.2021YFA0718500) from the Minister of Science and Technology of China (MOST). The authors thank supports from the National Natural Science Foundation of China under Grants 12273043, U1838201, U1838202, U1938102, and U1938108. This work was partially supported by International Partnership Program of Chinese Academy of Sciences (Grant No.113111KYSB20190020).

\bibliography{sn-article}% common bib file

\end{document}